\documentclass[aps,twocolumn,pra,superscriptaddress,showpacs,tightenlines]{revtex4}
\usepackage{mathrsfs}
\usepackage{amssymb}
\usepackage{amsmath}
\usepackage{graphicx}
\usepackage{epsfig}
\usepackage{txfonts}
\usepackage{subfigure}
\usepackage{amsfonts}

\begin{document}

\title{Controlling single-photon scattering in a rectangular waveguide by a
 V-type three-level emitter}

\author{Ya-Ju Song}
\affiliation{Graduate School of Chinese Academy of Engineering Physics, Beijing 10084, China}
\affiliation{Department of Physics, Hengyang Normal University, Hengyang 421002, China}
\author{Lei Qiao}
\affiliation{Graduate School of Chinese Academy of Engineering Physics, Beijing 10084, China}
\author{Chang-Pu Sun}
\thanks{Corresponding author}
\email{cpsun@csrc.ac.cn}
\affiliation{Graduate School of Chinese Academy of Engineering Physics, Beijing 10084, China}
\affiliation{Beijing Computational Science Research Center, Beijing 100193, China}

\date{\today}

\begin{abstract}
The single-photon scattering in a rectangular waveguide by a V-type three-level emitter is studied for large range of input-photon energy beyond the single-mode region. By using Lippmann-Schwinger formalism, the necessary and sufficient conditions of complete transmission and complete reflection are derived analytically. In the single-mode region, the complete transmission caused by electromagnetically induced transparency (EIT) and the complete reflection due to Fano resonance can both be achieved by adjusting the emitter's parameters.
But in the multi-mode region, except that the input-state is prepared in a coherent superposition state, the perfect reflection is absent, and the photon inevitably enters other propagation modes due to the indirectly interaction between waveguide modes mediated by the emitter. Other remarkable features in the photon transport induced by the finite cross section includes the blueshift of the reflection resonance and the cutoff-frequency effect.
\end{abstract}


\maketitle \narrowtext

\section{Introduction}

Waveguide quantum electrodynamics (QED) studies the interaction between the electromagnetic continuum confined to waveguides and quantum emitters (natural or artificial atoms). Owing to the integrable advantage superior to traditional cavity QED, waveguide QED has extensive applications in quantum information processing, such as quantum network \cite{Kimble2008}, quantum communication \cite{Hoi2011,Zhou2013,Tiecke2014,Shomroni2014} and quantum simulation \cite{Garcia2015}. There are a wide variety of promising candidates for waveguide QED systems, including Quantum dots coupled to a nanowire \cite{Claudon2010,Munsch2012,Huck2016,Akimov2007,Akselrod2014,Versteegh2014}, atoms embedded in an optical fiber\cite{Bajcsy2009,Babinec2010}, superconducting qubits with a superconducting transmission line resonator \cite{Hoi2011,Astafiev2010,Van Loo2013,Hoi2013,Hoi2012,Hoi2013b,Astafiev2010b,Abdumalikov2010}, and photonic crystal waveguides with quantum dots \cite{Arcari2014,Lund2008,Laucht2012} or natural atoms \cite{Goban2014}. In waveguide QED, it is a serious subject to control the photon transport in waveguides by interaction with quantum emitters. This study not only deepens the understanding of the fundamental but important light-matter interaction, but also facilitates the development of single-photon devices with different functions, like quantum switches\cite{Zhou2008}, optical transistors \cite{Chang2007}, optical routings \cite{Hoi2011,Zhou2013,Tiecke2014,Shomroni2014}, and frequency converters \cite{Wang2014}.

Over the past few years, the single-photon or few-photon scattering in one-dimensional (1D) waveguides by two-level emitters has been intensively studied by several approaches \cite{Roy2017,Liao2016,Shen2005,Shen2005b,Taylor1972,Schneider2016,Roy2011,Fan2010,Lalumiere2013,Xu2015,Shi2009,Rephaeli2010,Shi2015,Liao2016b,Greenberg2015,Shen2009,Zheng2010}, including Bethe-anatz approach \cite{Zhou2008,Shen2005,Shen2005b}, Lippmann-Schwinger formalism \cite{Taylor1972,Schneider2016,Roy2011,Qiao2017}, input-output theory \cite{Fan2010,Lalumiere2013,Xu2015}, path integral formalism \cite{Shi2009}, and dynamical theory \cite{Rephaeli2010,Shi2015,Liao2016b}. A key property in these work is that the photon will be perfect reflected when the incident photon energy resonates with the two-level emitter, due to destructive interference between the directly transmitted photon and the photon re-emitted by the emitter (Fano resonance). However, most of the work focuses on the 1D waveguide with linearized dispersion and does not consider the effect of cross section. A few studies have shown that the nonlinear dispersion relation of the 1D waveguide would lead to more interesting phenomena \cite{Zhou2008,Roy2011,Lombardo2014,Gonzalez2017,Alexanian2010}, like frequency shift of the reflection resonance and cutoff frequency effect. And the finite cross section would result in multiple mode effect, where the photon is inevitably scattered into other propagation modes \cite{Huang2013,Li2014,Song2018}. What is more, if the two-level emitters are replaced by three-level emitters \cite{Sanchez2016,Witthaut2010,Elfving2018,Elfving2018b,Zhou2007,Li2015,Gong2008}, the systems may exhibit richer behaviors such as electromagnetically induced transparency (EIT). Thus, a more challenging task is to study the photon transport property in the realistic waveguide with nonlinear dispersion relation and finite cross section.

In this letter we analyze the single photon scattering in a rectangular waveguide by a V-type three-level emitter, where the two upper states of emitter separately couple to the ground state via the rectangular waveguide. Here we mainly explore the regulation mechanism of the three-level emitter on the photon transport, and take into account the effect of finite rectangular cross section of the waveguide. To this end, we use the Lippmann-Schwinger equation to analytically derive the reflection and transmission amplitudes.
In the single mode case, where the incident photon energy is smaller than the second lowest cutoff frequency, the complete transmission and complete reflection can both be achieved due to the EIT and Fano resonance, respectively. However, with the increasing incident photon energy, the multi-mode effect induced by finite cross section can not be ignored. But it can be eliminated by setting the input-state in a certain coherent superposition state (CSS) \cite{Li2014}, such that the completely destructive interference between the directly transmitted photon and the re-emitted photon occurs under the Fano resonance condition. Here the re-emitted photon stems from the emitter's response corresponding to multiple scattering pathways with different propagation modes.

The paper is organized as follows: In Sec.~II, we introduce the Hamiltonian model. In Sec.~III, with the use of Lippmann-Schwinger equation, we derive the scattering matrix $S$ of the rectangular waveguide doped with the V-type three-level emitter. In Sec.~IV, we aim to explore the regulation mechanism of the three-level emitter on the photon transport, and take into account the effect of finite rectangular cross section of the waveguide. Finally, Sec.V is devoted to some conclusions.

\section{\label{Sec:2}Physical model}
\begin{figure}
\centering{\includegraphics[clip=true,height=4.5cm,width=7cm]{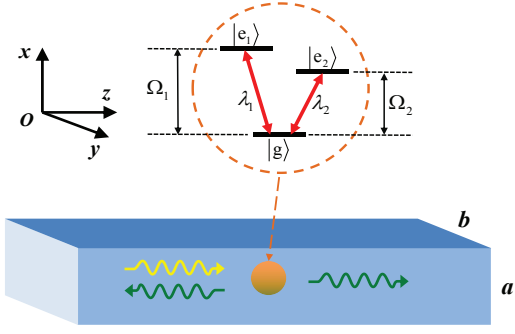}}
\caption{(Color online) Schematic of controlling single-photon scattering in an infinite rectangular waveguide by a V-type three-level emitter. The emitter is located in the center of cross section and dipole-coupled to the TM modes of the waveguide. Both upper states $\vert e_{1}\rangle $ and $\vert e_{2}\rangle$ couple to the waveguide mode with the strengths $\lambda _{1}$ and $\lambda _{2}$, respectively. The transition frequencies between two upper states and ground state $\vert g\rangle $ are $\Omega_{1}$ and $\Omega_{2}$.}\label{fig1}
\end{figure}
We consider an infinite rectangular waveguide interacts with a V-type three-level emitter, as shown in Fig.~\ref{fig1}. The emitter is located in the center of the cross section with the location $\vec{r}=(a/2,b/2,0)$. The $a$ and $b$ are the length and width of the cross section, respectively. Both two upper states $\vert e_{1}\rangle $ and $\vert e_{2}\rangle $ of the emitter are coupled to the ground state $\vert g\rangle$ via the rectangular waveguide. Assuming the emitter's electric dipole is oriented
along the longitudinal $z$ direction, the emitter only interacts with the transverse
magnet (TM) modes of the rectangular waveguide.
The TM modes are labeled by the longitudinal propagating wave number $k$ and two positive odd integers ($m, n$) with the standing wave
numbers in the cross section $k_{x}=m\pi /a$, $k_{y}=n\pi/b$. Hereafter the labels ($m, n$) are replaced by the sequence number $j$ according to the ascending order of the cutoff frequencies $\omega_{j}=c\pi\sqrt{( m/a) ^{2}+( n/b) ^{2}}$. In this paper, the cross sectionial parameters are set to be $a=1.5b$, thus $j = 1,2,3,...$ denote TM$_{11}$, TM$_{31}$, TM$_{13}$, TM$_{51}$... , respectively. Note that the ratio of $a$ to $b$ determines the order of different TM modes.

The Hamiltonian for the total system consists of the free and interaction parts. $H_{w}$ and $H_{e}$ denote the free Hamiltonian of waveguide and emitter, respectively,
\begin{eqnarray}
\label{eq-1}
H=H_{0}+V,\ \ \ H_{0}=H_{w}+H_{e}.
\end{eqnarray}

Firstly, according to the Maxwell equations with the boundary condition of rectangular waveguides, the TM-mode waveguide field can be described by the vector potential $(\hbar =c=\varepsilon _{0}=1)$,
\begin{eqnarray}
\vec{A}=\sum_{j}\int_{-\infty }^{+\infty }dk\frac{1}{\sqrt{\pi
ab\omega _{j,k}}}\left[ \vec{\mu}_{j,k}\left( \vec{r}\right)
a_{j,k}\left( t\right) +H.c.\right],
\end{eqnarray}
with the mode functions being $\vec{\mu}_{j,k}^{(x)} =\frac{kk_{x}e^{{}ikz}}{\omega_{j}\omega _{k}}\cos( k_{x}x) \sin( k_{y}y)$, $ \vec{\mu}_{j,k}^{( y)} =\frac{kk_{y}e^{{}ikz}}{\omega_{j}\omega _{k}}\sin ( k_{x}x) \cos ( k_{y}y)$, $\vec{\mu}_{j,k}^{(z) } =\frac{-i\omega _{j}e^{{}ikz}}{\omega_{k}}\sin ( k_{x}x) \sin ( k_{y}y)$. The $a_{j,k}^{\dag}$ ($a_{j,k}$) denotes the photon creation (annihilation) operator for the TM$_{mn}$ mode with the longitudinal wave number $k$, $a_{j,k}(t)= a_{j,k}e^{-i\omega _{j,k}t}$, and the $\omega _{j,k}$ represents the
nonlinear dispersion relation of the rectangular waveguide,
\begin{eqnarray}
\label{eq-3}
\omega _{j,k}&=&\sqrt{\omega _{j}^{2}+k^{2}}.
\end{eqnarray}
Then the electric and magnetic fields can be derived from $\vec{E} =-\frac{\partial \vec{A}}{\partial t}-\nabla\varphi$ and $\vec{B}=\nabla \times \vec{A}$ with the Coulomb gauge ($\nabla \cdot \vec{A} =0$ and $\varphi=0$). And the Hamiltonian of rectangular waveguide can be derived from $H_{w}=\frac{1}{2}\int d\vec{r}\left( \vec{E}^{2}+\vec{B}^{2}\right)$, which can be expressed as following after discarding the constant term,
\begin{eqnarray}
H_{w} &=&\sum_{j}\int_{-\infty }^{+\infty }dk \omega _{j,k} a_{j,k}^{\dagger
}a_{j,k}.
\end{eqnarray}

Furthermore, we adopt a three-level approximation for the emitter with the ground state $\vert g\rangle$ and two excited state $\vert e_{i}\rangle$ satisfying $H_{e}\vert g\rangle=0$ and $H_{e}\vert e_{i}\rangle=\Omega_{i}\vert e_{i}\rangle$. Thus, the Hamiltonian for the emitter is
\begin{eqnarray}
H_{e}&=&\sum\limits_{i=1,2}\Omega _{i}\vert e_{i}\rangle
\langle e_{i}\vert.
\end{eqnarray}

At last, the waveguide-emitter interaction Hamiltonian can be obtained by the minimal coupling prescription, $V=-\frac{q}{m}\vec{A}\cdot \vec{p}=-\frac{q}{m}A_{z}p_{z}$. Here we have adopted the long-wavelength and the electric dipolar approximations.
The matrix elements of emitter's momentum can be derived as $\left\langle e_{i}\right\vert p_{z}\left\vert g\right\rangle =i\Omega
_{i}m\left\langle e_{i}\right\vert z\left\vert g\right\rangle$ based on the canonical commutation relation $\vec{p}=-im[\vec{r},H_{e}]$. Combined with the component of $\vec{A}$ in the $z$ direction, the interaction Hamiltonian under the rotating wave approximation can be derived as
\begin{eqnarray}
\label{eq-2}
V&=&\sum\limits_{i=1,2}\sum_{j}\int_{-\infty }^{+\infty }dk\left(
g_{j,k}^{( i) }\vert e_{i}\rangle \langle
g\vert a_{j,k}+H.c.\right).
\end{eqnarray}
In Eq.~(\ref{eq-2}) the sum over $i=1,2$ includes two dipole-allowed transitions between the two upper states and the ground, and the sum over $j$
denote the interaction between the emitter and different TM modes.
Here the $g_{j,k}^{(i)}$ is mode-dependent coupling strength,
\begin{eqnarray}
\label{eq-4}
g_{j,k}^{(i)}&=&-\frac{\lambda _{i}\Omega_{i}\omega _{j}}{\omega _{j,k}^{3/2}}\sin
\left( \frac{m\pi }{2}\right) \sin \left( \frac{n\pi }{2}\right),
\end{eqnarray}
where the dimensionless coupling strength $\lambda _{i}=p_{i}/\sqrt{\pi ab}$ is determined by the size of cross section and the emitter's electric moment $p_{i}=q\langle e_{i}\vert z \vert g\rangle$.
\section{\label{Sec:2}Scattering matrix $S$}
We now derive the scattering matrix $S$ of the rectangular waveguide doped with the V-type three-level emitter.
For a single-photon input state $\vert \varphi _{in}\rangle =\vert \varphi_{j,k}\rangle$ with energy $E=\omega _{j,k}$, the scattering state in the single-excitation subspace can be expressed as
\begin{eqnarray}
\label{eq-5}
\vert \varphi _{j,k}^{(+)}\rangle  = \sum_{j^{\prime
}}\int_{-\infty }^{+\infty }dk^{\prime }u(j^{\prime },k^{\prime
};j,k)\vert \varphi _{j^{\prime },k^{\prime }}\rangle +\sum\limits_{i=1,2}u_{e}^{(i)
}( j,k)\vert e_{i}\rangle.
\end{eqnarray}
Here the $\vert \varphi_{j,k}\rangle$ denotes the waveguide in the $j$-th mode single-photon state and the emitter in the ground state, while the $\vert e_{i}\rangle$ denotes the waveguide field in vacuum and the emitter in the $i$th upper state.

Submitting Eqs.~(\ref{eq-1})-(\ref{eq-5}) into the Lippmann-Schwiger equation \cite{Taylor1972},
\begin{eqnarray}
\label{eq-6}
\vert \varphi _{j,k}^{(+)}\rangle &=&\vert \varphi
_{j,k}\rangle +\frac{1}{E-H_{0}+i0^{+}}V\vert \varphi
_{j,k}^{(+)}\rangle,
\end{eqnarray}
we can get the amplitude components of the scattering state,
\begin{eqnarray}
\label{eq-7}
&u_{e}^{( 1) }( j,k)= \frac{g_{j,k}^{(1)
}( E-\Omega _{2}) }{f( E) },\ \ \ \
u_{e}^{( 2) }( j,k)= \frac{g_{j,k}^{(2)
}( E-\Omega _{1}) }{f(E) }, \nonumber\\
&u(j^{\prime },k^{\prime };j,k)= \frac{\sum\limits_{q=1,2}g_{j^{\prime
},k^{\prime }}^{\ast (q) }u_{e}^{(q) }(
j,k) }{E-\omega _{j^{\prime },k^{\prime }}+i0^{+}}+\delta _{j^{\prime
},j}\delta( k^{\prime }-k),
\end{eqnarray}
with the function $f(E) $ defined by
\begin{eqnarray}
\label{eq-8}
f(E)&=&( E-\Omega _{1})( E-\Omega _{2})-( E-\Omega _{2})h^{(1)}(E)\nonumber\\
&&-( E-\Omega _{1})h^{(2)}(E).
\end{eqnarray}
Note that the $h^{(i)}(E)$ $(i=1,2)$ is determined jointly by the coupling strength in Eq.~(\ref{eq-4}) and the nonlinear dispersion relation in Eq.~(\ref{eq-3}),
\begin{eqnarray}
\label{eq-8}
h^{(i)}(E)=\sum_{j^{\prime }}\int_{-\infty }^{+\infty } dk^{\prime }\frac{\left|g_{j,k}^{(i)}\right|^{2}}{E-\omega _{j^{\prime },k^{\prime
}}+i0^{+}}.
\end{eqnarray}
Using integral formulas and the property $\lim_{\varepsilon \rightarrow 0^{+}}\frac{1}{%
x-x_{0}+i\varepsilon }=\frac{\mathcal{P}}{x-x_{0}}-i\pi \delta \left(
x-x_{0}\right) $ with $\mathcal{P}$ being the principal value function, we can obtain
the real and imaginary parts of the $h^{(i)}\left( E\right) =\Delta^{(i)}
\left( E\right) -i\Gamma^{(i)} \left( E\right) $. The $\Delta^{(i)}(E)=\sum_{j}\Delta^{(i)}_{j}(E)$, $\Gamma^{(i)}( E)=\sum_{j}\Gamma^{(i)}_{j}(E)$ consist of the contribution of TM modes with
\begin{eqnarray}
\label{eq-10}
\Delta^{(i)}_{j}(E)
&=&\frac{\Omega_{i}^{2}|\lambda_{i}|^{2}}{E^{2}}\left[2E+\pi\omega_{j}%
+\frac{\omega_{j}^{2}\zeta_{j}(E)}{\sqrt{|\omega_{j}^{2}-E^{2}|}}\right],\nonumber\\
\Gamma^{(i)}_{j}( E)&=&\frac{2\pi \Omega_{i}^{2}|\lambda_{i}|^{2}\omega _{j}^{2}}{E^{2}\sqrt{E^{2}-\omega _{j}^{2}}}\Theta (
E-\omega _{j}).
\end{eqnarray}
Here the subscripts denote different waveguide modes, while the superscripts with brackets denote the emitter's two excited states. When $E>\omega _{j}$, $\zeta_{j}(E)=2\ln(\frac{\omega_{j}}{E-\sqrt{E^{2}-\omega_{j}^{2}}})$ leads to blue Lamb shift, while when $E<\omega _{j}$, $\zeta_{j}(E)=-\pi-2\arctan(\frac{E}{\sqrt{\omega_{j}^{2}-E^2}})$ leads to red Lamb shift. As a result of the weaker coupling and larger detuning with higher-frequency TM modes, the red Lamb shift is neglected in the following discussion. Eq.~(\ref{eq-10}) also indicates that, due to the limitation of the step function $\Theta (E-\omega_{j})$, the decay rate $\Gamma^{(i)}_{j}(E) $ will vanish if the incident-photon energy is smaller than the cutoff frequency.

Then the scattering matrix elements can be obtained according to the scattering states,
\begin{eqnarray}
\label{eq-11}
\langle \varphi _{j^{\prime },k^{\prime }}\vert S \vert
\varphi _{j,k}\rangle
&=&\delta _{j^{\prime },j}\delta( k^{\prime }-k) -2\pi i\delta
( \omega _{j^{\prime },k^{\prime }}-\omega _{j,k})\nonumber\\
&&\times \sum\limits_{i=1,2}g_{j^{\prime },k^{\prime }}^{\ast (i)
}u_{e}^{(i)}(j,k),
\end{eqnarray}
where the $u_{e}^{(i) }( j,k)$ is given by Eqs.~(\ref{eq-7})-(\ref{eq-10}). The first term on the right hand side of Eq.~(\ref{eq-11}) represents the directly transmitted photon, and the second term reflects the scattering process of the three-level emitter to the single photon. The delta function limits the photon to scattering only into the TM modes of energy conservation.
\section{\label{Sec:3} Single-photon scattering by a V-type emitter with finite cross section}
In this section, we explore the regulation mechanism of a V-type three-level emitter on the single-photon transport in the rectangular waveguide. The influence of the input-state and the multi-mode effect on the photon transport is also analyzed.

Without loss of generality, assume the input photon with the energy $E=\omega_{in}$ is in the superposition state,
\begin{eqnarray}
\label{eq-12}
\vert \varphi _{in}\rangle &=& \sum_{j=1}^{j_{\max}}c_{j}\vert \varphi_{j,k_{j}}\rangle,
\end{eqnarray}
where the number of TM modes $j_{\max }$ involved in photon scattering process is determined by the inequality $\omega _{j_{\max }}\leq \omega_{in}<\omega _{j_{\max }+1}$, $c_{j}$ is the normalized probability amplitude in the $j$th TM mode, and $k_{j}=\sqrt{\omega_{in}^{2}-\omega_{j}^{2}}$ is the corresponding wave vector for the given energy $\omega_{in}$.

Then the outgoing state can be derived via the scattering matrix $S$ in Eq.~(\ref{eq-11}),
\begin{eqnarray}
\label{eq-13}
\vert \varphi _{out}\rangle &=& S\vert \varphi
_{in}\rangle=\sum_{j=1}^{j_{\max }}\left[ (c_{j}+r_{j})\vert \varphi
_{j,k_{j}}\rangle +r_{j}\vert \varphi _{j,-k_{j}}\rangle
\right],
\end{eqnarray}
with the coefficients $r_{j}$ defined by
\begin{eqnarray}
\label{eq-14}
r_{j}&=&-2\pi i\rho _{j}\sum_{j^{\prime }=1}^{j_{\max
}^{\prime }}\sum\limits_{i=1,2}c_{j^{\prime
}}g_{j,k_{j}}^{\ast \left( i\right) }u_{e}^{\left( i\right) }\left(
j^{\prime },k_{j^{\prime }}\right).
\end{eqnarray}
During the calculation, we have used the property $\delta ( \omega
_{j,k}-\omega_{in}) =\rho_{j}[\delta (k-k_{j}) +\delta(k+k_{j})]$ with $\rho_{j}=\omega_{in}/\sqrt{\omega_{in}^{2}-\omega_{j}^{2}}$ being the photon state density in
the $j$th TM mode.

Therefore, the reflectivity and transmissivity in the $j$th TM
mode can be derived by submitting the Eq.~(\ref{eq-14}) into the definition
\begin{eqnarray}
\label{eq-15}
R_{j}=\frac{\vert r_{j}\vert^{2}/\rho_{j}}{\sum_{j^{\prime }=1}^{j_{\max
}^{\prime }}\vert c_{j^{\prime }}\vert^{2}/\rho_{j^{\prime }}},\ \ \
T_{j}=\frac{\vert c_{j}+r_{j}\vert^{2}/\rho_{j}}{\sum_{j^{\prime }=1}^{j_{\max
}^{\prime }}\vert c_{j^{\prime }}\vert^{2}/\rho_{j^{\prime }}}.
\end{eqnarray}
The total reflectivity and transmissivity are $R=\sum_{j=1}^{j_{\max }}R_{j}$ and $T=\sum_{j=1}^{j_{\max }}T_{j}$. Combined with the expressions of $g_{j,k}^{i}$ and $u_{e}^{i}(j,k)$ in Eqs.~(\ref{eq-4}) and~(\ref{eq-7}), it is easily to prove that $R=1-T$ and
\begin{eqnarray}
\label{eq-16}
R&=&\frac{\left|\frac{\text{Im}[{f(\omega_{in})} ]}{f(\omega_{in})}\right|^{2}\vert\sum_{j^{\prime }=1}^{j_{\max
}^{\prime }}c_{j^{\prime }}\omega_{j^{\prime }}\vert^{2}}{\sum_{j^{\prime }=1}^{j_{\max
}^{\prime }}\frac{\omega_{j^{\prime}}^{2}}{\sqrt{E^{2}-\omega_{j^{\prime}}^{2}}}\sum_{j^{\prime }=1}^{j_{\max
}^{\prime }}\vert c_{j^{\prime }}\vert^{2}\sqrt{E^{2}-\omega_{j^{\prime}}^{2}}}.
\end{eqnarray}
Eq.~(\ref{eq-16}) indicates that the single-photon transport in the rectangular waveguide can be manipulated not only by the input-state parameters $c_{j}$, but also by the emitter parameters $\lambda_{i}$ and $\Omega_{i}$ through the function $f(\omega_{in})$.

Noting that two special cases should be pointed out. In the degenerate upper states case with $\Omega_{1}=\Omega_{2}=\Omega$ and the two-level emitter case with $\lambda_{2}=0$, the $f(\omega_{in})$ in Eq.~(\ref{eq-16}) should be replaced as follows:
\begin{eqnarray}
\label{eq-17}
&&f(\omega_{in})\stackrel{\Omega_{1}=\Omega_{2}}\longrightarrow  \omega_{in}-\Omega-h^{(1)}(\omega_{in})-h^{(2)}(\omega_{in}), \nonumber\\
&&f(\omega_{in})\stackrel{\lambda_{2}=0}\longrightarrow \omega_{in}-\Omega_{1}- h^{(1)}(\omega_{in}).
\end{eqnarray}

\subsection{Perfect transmission and perfect reflection conditions}
Now we begin to analyze the conditions for perfect transmission and perfect reflection from Eq.~(\ref{eq-16}). Obviously, the photon is completely transmitted with  $\left\vert\varphi_{out}\right\rangle =\left\vert \varphi _{in}\right\rangle$ under either of the following two conditions,
\begin{eqnarray}
\label{eq-18}
T=1:\ \ \sum_{j=1}^{j_{max}}c_{j}\omega_{j}=0\ \ or\ \ \text{Im}[f(\omega_{in})]=0.
\end{eqnarray}
The first condition is the requirement for the input-state parameters.
As long as the superposition coefficient of the input-state satisfies $\sum_{\emph{j}^=1}^{\emph{j}_{max}}\emph{c}_{\emph{j}}\omega_{\emph{j}}=0$, the three-level
emitter has no effect on the single-photon transport no matter what the system parameters are. The other condition for complete transmission means that, by controlling the emitter's parameters ($\lambda_{i}$ and $\Omega_{i}$) to satisfy the condition of $\text{Im}[f(\omega_{in})]\propto y(\omega_{in})=\vert \lambda _{1}\vert ^{2}\Omega_{1}^{2}( \omega_{in}-\Omega
_{2}) +\vert \lambda _{2}\vert ^{2}\Omega_{2}^{2}( \omega_{in}-\Omega
_{1})=0$, the complete transmission can be achieved for any input states in Eq.~(\ref{eq-12}). Noting that by using a two-level emitter or a degenerate three-level emitter, the complete transmission can not occur as the $f(E)$ in Eq.~(\ref{eq-16}) should be replaced as in Eq.~(\ref{eq-17}). This exactly highlights the advantages of non-degenerate three-level emitter.

The physical mechanism of complete transmission can be explained as a result of destructive interference. Returning to the Eq.~(\ref{eq-14}), we find there exist two kinds of interference in the scattering process. The first one corresponds to the sum of $j$: when the incident photon is set to be the superposition state in Eq.~(\ref{eq-12}), there appear multiple scattering channels from $\vert \varphi_{j,k_{j}}\rangle$ to $\vert e_{i}\rangle$. When the superposition coefficient meets the condition of $\sum_{{j}^=1}^{{j}_{max}}{c}_{{j}}\omega_{{j}}=0$ derived from $\sum_{{j}^=1}^{{j}_{max}}{c}_{{j}}g_{{j,k_{j}}}^{(i)}=0$ (dark states), both transitions are suppressed due to destructive interference, and the emitter becomes transparent with respect to the input photon.
The second kind of interference corresponds to the sum of $i$: the excited emitter after absorbing the photon has two transition pathways to the ground state. Under the condition of $y(\omega_{in})=0$, the destructive interference between these two radiation pathways leads to the perfect transmission, the so-called EIT mechanism. But for the degenerate three-level emitter, the photon from these two radiation pathways cannot be distinguished, thus the destructive interference phenomenon vanishes.

We further analyze the conditions for perfect reflection without considering the cutoff-frequency effect due to $\lim_{\omega_{in}\rightarrow\omega_{j}}\rho_{j}\rightarrow \infty$ (zero group velocity). With the use of Cauchy-Buniakowsky-Schwarz inequality, we can find the conditions for perfect reflection,
\begin{eqnarray}
\label{eq-20}
R=1:\left(j_{\max}=1\ or\ c_{j^{\prime }}\propto \frac{\omega _{j^{\prime }}}{\sqrt{E^{2}-\omega _{_{j^{\prime }}}^{2}}}\right)\ and\ \text{Re}[f(\omega_{in})]=0.\nonumber\\
\end{eqnarray}
There exist two kinds of special input-state for perfect reflection. The first one is that the input-photon energy in the region of $\omega_{1}< \omega_{in}<\omega_{2}$ ($j_{\rm{max}}=1$) limits only the TM$_{11}$ mode to involve in scattering process. The second one is that the input-photon is in multiple modes with the superposition coefficients restricted by $c_{j^{\prime }}\propto \omega _{j^{\prime }}/\sqrt{\omega_{in}^{2}-\omega _{_{j^{\prime }}}^{2}}$, we call coherent superposition states (CSS).
Only if the input-state is in the single-mode region or prepared in CSS can the Eq.~(\ref{eq-14}) be simplified to $r_{j}=-c_{j}\text{Im}[{f(\omega_{in})} ]/f(\omega_{in})$. This is actually the result of the coherent interference between different scattering channels from $\vert \varphi_{j,k_{j}}\rangle$ to $\vert e_{i}\rangle$. That is to say, even in the multi-mode region, the effect of emitter on the photon scattering in each TM mode is the same as long as the input state is the CSS. Then the photon scattering with CSS input-state is similar to the single-mode case. The total reflectivity with two special input states can both be reduced to
\begin{eqnarray}
\label{eq-22}
R &=&\frac{\text{Im}[{f(\omega_{in})} ]^{2}}{|f(\omega_{in})|^{2}}.
\end{eqnarray}
The only difference is that $\Gamma^{(i)}(\omega_{in})=\Gamma^{(i)}_{1}(\omega_{in})$ for the single-mode case, while $\Gamma^{(i)}(\omega_{in})=\sum_{j=1}^{j_{\max}}\Gamma^{(i)}_{j}(\omega_{in})$ for the CSS case.

The other necessary condition of perfect reflection is requirement for the emitter's parameters to meet $\text{Re}[f(\omega_{in})]\equiv(\omega_{in}-\Omega _{1})(\omega_{in}-\Omega_{2}) -(\omega_{in}-\Omega_{2})\Delta^{(1)}(\omega_{in})-(\omega_{in}-\Omega_{1})\Delta^{(2)}(\omega_{in})=0$. It forecasts that the energy of perfect reflected photon resonates with the re-normalized transition frequency between $\vert e_{i}\rangle$ and $\vert g\rangle$. The frequency shift of the resonant frequency with respect to the bare frequency $\Omega_{i}$ is determined by $\Delta^{(i)}(\omega_{in})$, which stems from the finite cross section and the nonlinear dispersion relation of the rectangular waveguide. Then two different input-photon energy for complete reflection could reasonably be expected to exist. But for the degenerate three-level emitter or the two-level emitter cases, there exists at most one input-photon energy for complete reflection, which can be derived from Eq.~(\ref{eq-17}). It is worth noting that
the complete reflection can be explained as a result of Fano interference between the directly transmitted photon and the reemitted photon. This can be proved by the probability amplitude of transmitted photon in the outgoing state in Eq.~(\ref{eq-13}), i.e., $c_{j}+r_{j}$. Only if the resonance condition with $\text{Re}[f(\omega_{in})]=0$ and either of two special input states with $r_{j}=-c_{j}\text{Im}[f(\omega_{in})]/f(\omega_{in})$ are simultaneously satisfied, the directly transmitted photon and the photon re-emitted by the emitter only differ by a phase shift $\pi$. Then the perfect reflection phenomenon occurs, as a result of completely destructive interference of the transmitted photon amplitudes.
\subsection{Controlling single-photon transport by a three-level emitter in single-mode region}
\begin{figure}
\centering
\subfigure{\includegraphics[clip=true,height=4.5cm,width=7.0cm]{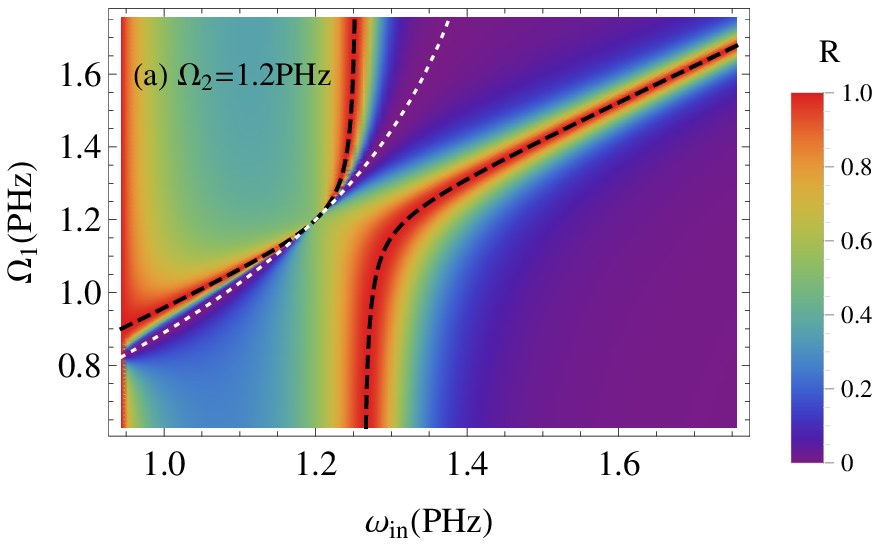}}
\subfigure{\includegraphics[clip=true,height=4.3cm,width=7.0cm]{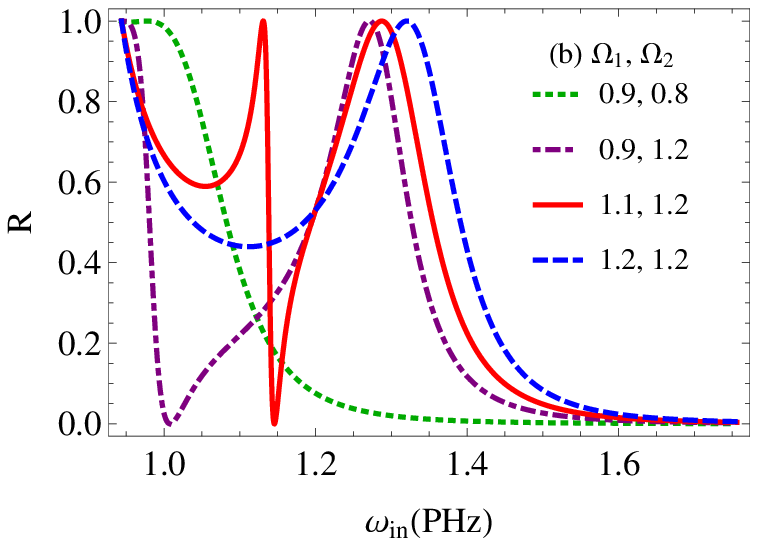}}
\subfigure{\includegraphics[clip=true,height=4.3cm,width=7.0cm]{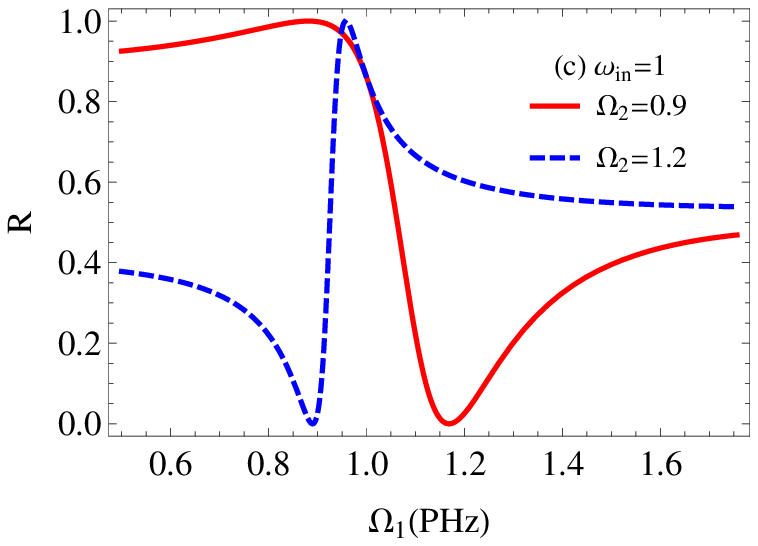}}
\caption{(Color online)  Single-mode region: (a) Density plot of reflectivity $R$ as a function of   input-photon energy $\omega_{1}<\omega_{in}<\omega_{2}$ and the transition frequency $\Omega_{1}$ with $\Omega_{2}=1.2$. The dotted (white) and dashed (black) curves correspond to complete transition and complete reflection, respectively. (b) $R$ against $\omega_{in}$ for different values of transition frequencies.
 (c) $R$ against $\Omega_{1}$ for a given input-photon energy $\omega_{in}=1$ with different $\Omega_{2}=0.9,1.2$.
Other parameters are chosen as $\lambda_{1}=\lambda_{2}=0.1$, $b=1.2$ um, $a=1.5b$. All frequencies are in units of the PHz.}\label{fig3}
\end{figure}
\begin{figure}
\centering
\subfigure{\includegraphics[clip=true,height=4.3cm,width=7.0cm]{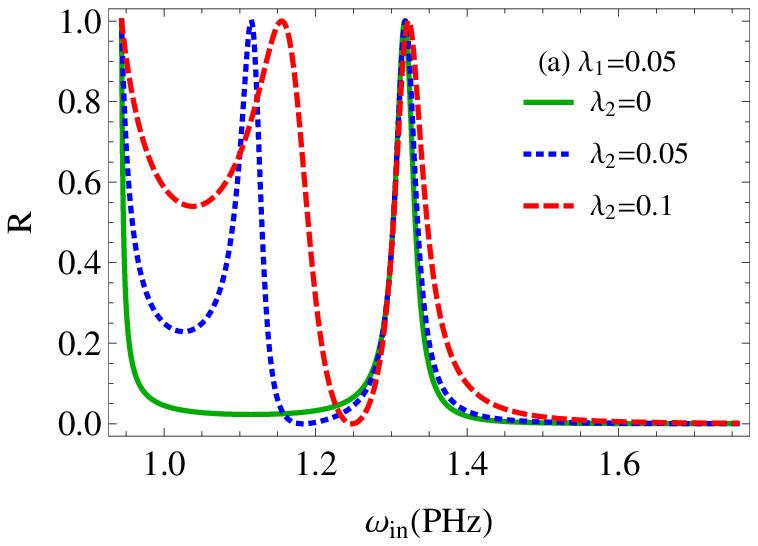}}
\subfigure{\includegraphics[clip=true,height=4.3cm,width=7.0cm]{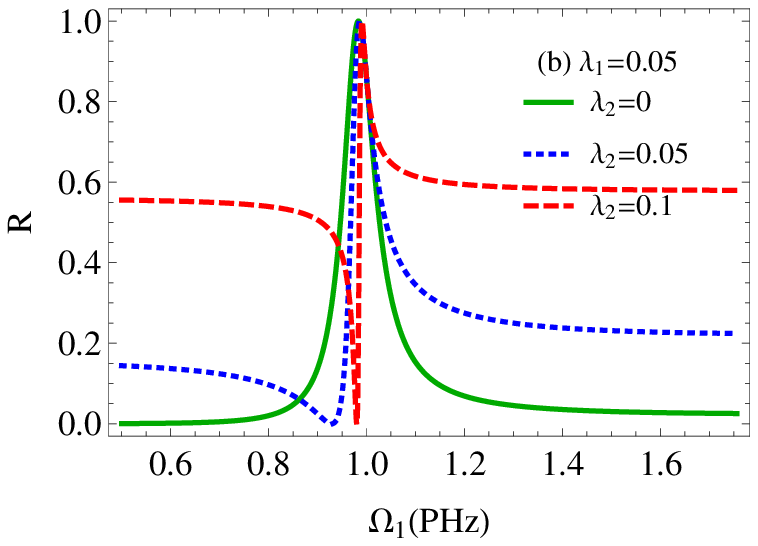}}
\caption{(Color online) Single-mode region: For different dimensionless coupling strengths $\lambda_{i}$ caused by the change in electric dipole moments, (a) $R$ as a function of $\omega_{in}$ with $\Omega_{1}=1.3$ and $\Omega_{2}=1.1$. (b) $R$ against $\Omega_{1}$ for a given input-photon energy $\omega_{in}=1$ with $\Omega_{2}=1.1$. The size of cross section is $b=1.2$ um, $a=1.5b$. All frequencies are in units of the PHz.}\label{fig4}
\end{figure}

In order to explore how the photon transport is manipulated by the V-type three-level emitter, we further investigate the effect of emitter's parameters on the reflectance spectrum in the simple single-mode case ($j_{\max}=1$). Note that only when the photon frequency is higher than the cutoff frequency of dominant TM$_{11}$ mode, but lower than the second lowest cutoff frequency, i.e., $\omega_{1}<\omega_{in}<\omega_{2}$, the waveguide works in the single-mode region. In the single-mode region, only the TM$_{11}$ mode interacts with the emitter. Here we also assume $\Omega_{i}<\omega_{2}$ to ignore the multi-mode effect. Note that the photon will be reflected completely when the energy resonates with the cutoff-frequency, i.e., $R(\omega_{in}=\omega_{1})=1$, as a result of $\lim_{\omega_{in}\rightarrow\omega_{1}}\rho_{1}\rightarrow \infty$.

Firstly, the effect of the $\Omega_{i}$ on the photon transport is illustrated in Fig.~\ref{fig3} with $\lambda_{1}=\lambda_{2}=0.1$. As shown in Fig.~\ref{fig3}(a), the dotted white and dashed black lines hint that the conditions for perfect transition and perfect reflection are exactly correspond to $\text{Im}[f(\omega_{in})]=0$ and $\text{Re}[f(\omega_{in})]=0$ , respectively. Here the size of cross section is set to be $b=1.2$um, $a=1.5b$, with the cutoff frequency $\omega_{1}=0.94$PHz, $\omega_{2}=1.75$PHz \cite{Bajcsy2009}. In the weak-coupling limit, the energy of photon completely reflected is approximately the transition frequency $\omega_{in}\approx\Omega_{i}$. While the condition for perfect transition reduces to $\omega_{in}=\frac{\Omega_{1}^{2}\Omega_{2}+\Omega_{1}\Omega_{2}^{2}}{\Omega_{1}^{2}+\Omega_{2}^{2}}$ when $\lambda_{1}=\lambda_{2}$. Looking more closely, we find that the transmission spectral width (blue region) increases with the frequency difference $\vert\Omega_{1}-\Omega_{2}\vert$. In the region of $\Omega_{1}<\Omega_{2}$, the reflection spectral width (red region) around $\Omega_{1}$ decreases with $\Omega_{1}$, while in the region of $\Omega_{1}>\Omega_{2}$, the opposite is true.

Fig.~\ref{fig3}(b) shows that the detailed behavior of the reflectance spectrum primarily depends on the number of solutions to the two condition equations:
(i) When $\Omega_{i}<\omega_{1}$ ($i=1,2$), both two condition functions have no solutions in the region of $\omega_{1}<\omega_{in}<\omega_{2}$. With the increase of $\omega_{in}$, the detuninig between photon and emitter also increases gradually, and the scattering effect of emitter on photon weakens, showing a monotonic decline of reflectivity (dotted green line).
(ii) When $\Omega_{1}<\omega_{1}<\Omega_{2}<\omega_{2}$, the perfect reflection occurs around $\omega_{in}\approx\Omega_{2}$ as a result of Fano resonance, and the complete transmission similar to EIT may also be achieved if $\Omega_{i}$ are large enough to satisfy $\text{Im}[f(\omega_{in})]=0$. Thus, $R$  first gets smaller and then goes up to perfect reflection and finally decreases again (dot-dashed purple line).
(iii) When $\omega_{1}<\Omega_{i}<\omega_{2}$ ($i=1,2$), there exist two $\omega_{in}$ satisfying the Fano resonance condition and one $\omega_{in}$ satisfying the EIT condition. As a result, the $R$ exhibits two complete reflection peaks and one complete transmission valley (solid red line).
(iv) In the special degenerate three-level case with $\Omega_{1}=\Omega_{2}$, the $R$ exhibits only one perfect reflection peak and the perfect transition is absent (dashed blue line). This is due to the fact that the $f(\omega_{in})$ in this degenerate case should be replaced as in Eq.~(\ref{eq-17}).

For a given input-photon energy $\omega_{in}=1$PHz, the reflectivity $R$ as a function of the transition frequencies $\Omega_{i}$ is shown in Fig.~\ref{fig3}(c). The photon transport can be co-regulated by two transition frequencies. When $\Omega_{2}<\omega_{in}$, the $R$ increases to perfect reflection first, then decreases to perfect transmission and finally increases with $\Omega_{1}$. But when $\Omega_{2}>\omega_{in}$, the change of $R$ with respect to $\Omega_{1}$  is just the opposite.
Compared with the two-level emitter case, the regulation means is more flexible, which is beneficial to experimentally realize single-photon devices such as optical switches.

Secondly, we start to analyze the dependence of photon transport on the dimensionless coupling strength $\lambda_{i}$ in Fig.~\ref{fig4}. The $\lambda _{i}= p_{i}/\sqrt{\pi ab}$ is jointly decided by the emitter's electric dipole moments $p_{i}$ and the size of cross section ($a, b$). Here we only discuss the change of $\lambda_{i}$ caused by the electric moments, and the change caused by the size of cross section will be discussed in the next subsection.
In Fig.~\ref{fig4}(a) with $\omega_{1}<\Omega_{i}<\omega_{2}$ ($i=1,2$), the reflection $R$ exhibits two complete reflection peaks and one complete transition valley no matter what the nonzero coupling strengths are. This is consistent with the previous discussion on Fig.~\ref{fig3}: the reflectance spectrum primarily depends on the $\Omega_{i}$ in the weak-coupling regimes. And by comparing the dotted blue and the dashed red lines in Fig.~\ref{fig4}(a), it is found that, for a given $\lambda_{1}$, the reflection frequency and the reflection spectral width around the $\Omega_{2}$ increases with $\lambda_{2}$, while the transmission frequency changes in the direction of $\Omega_{1}$ and the transmission spectral width decreases with increasing $\lambda_{2}$.
This phenomenon can be explained as the blue shift of emitter's transition frequency increases with the coupling strength, as shown in Eq.~(\ref{eq-10}).
What is more, in the special case with $\lambda_{2}=0$, the V-type emitter reduces to a two-level emitter. The EIT-like phenomena induced by destructive interference between two radiation pathways disappears, and the single-photon can only be reflected completely under the Fano resonance condition (solid green line in Fig.~\ref{fig4}(a)).

Besides, for a given $\omega_{in}$ in the single-mode region, the regulatory capacity of two transition frequencies $\Omega_{i}$ on photon transport is determined by the coupling strengths $\lambda_{i}$. As shown in Fig.~\ref{fig4}(b), in the extreme case of $\lambda_{2}=0$, the photon transport characteristics are completely controlled by $\Omega_{1}$: under the weak coupling condition with $\lambda_{1}=0.05$, the two-level emitter is almost transparent to photons, unless the Fano resonance occurs.
But as the $\lambda_{2}$ is increased to be comparable or even greater than $\lambda_{1}$, the effect of transition between $|e_{2}\rangle$ and $|g\rangle$ on the photon transport cannot be ignored: it not only induces the EIT-like phenomenon, but also increases the sensitivity of the reflectivity $R$ to $\Omega_{1}$, as shown in the dotted blue and dashed red lines in Fig.~\ref{fig4}(b). This means that with the help of the third-level $|e_{2}\rangle$, it is possible to realize the regulation of photon transport from perfect transmission to perfect reflection within a small range of $\Omega_{1}$.

Until now, we can safely conclude that the V-type three-level emitter can be used to regulate the single-photon transport from complete transmission to complete reflection in the single mode region. Compared with the two-level emitter case, in which the complete transmission can not be achieved, there are more tunable parameters in the V-type three-level emitter case.

\subsection{Multi-mode effect induced by finite cross section}
\begin{figure}
\centering{\includegraphics[clip=true,height=5cm,width=8cm]{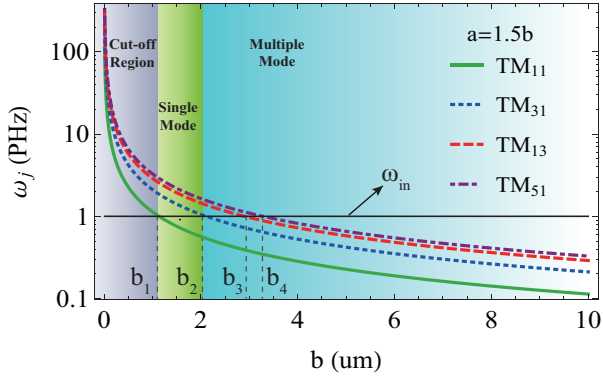}}
\caption{(Color online) The cutoff frequencies $\omega_{j}$ against the size of cross section. For a given input-photon energy $\omega_{in}$, there exist cut-off region, single-mode region, and multiple-mode region. The critical size $b_{j}$ is determined by $\omega_{in}=\omega_{j}$.}\label{fig2}
\end{figure}

We finish with a detailed study of the influence of multi-mode effect on the single-photon scattering in the rectangular waveguide. Recall that the single-mode region requires the photon energy to be in the region $\omega_{1}<\omega_{in}<\omega_{2}$. As shown in Fig.~\ref{fig2}, although the cutoff frequencies decrease with the size of cross section, the single-mode bandwidth $\omega_{2}-\omega_{1}$ becomes narrow (Note the vertical axis in Fig.~\ref{fig2} is logarithmic). For this reason, in most cases more than one waveguide mode is involved in photon scattering, and the multi-mode effect should be considered. Besides, driven by the development of high-frequency laser (like x-ray laser), the research on multi-mode effect becomes more urgent as the cross-section of realistic waveguide cannot be infinitely small.

Looking in closer detail expose that, for a given input-photon energy $\omega_{in}$, the number of TM modes involved in photon scattering is determined by the size of the waveguide's cross section. As shown in the Fig.~\ref{fig2}, the critical size $b_{j}=\frac{c\pi}{\omega_{in}}\sqrt{\frac{m^{2}}{l^{2}}+n^{2}}$ is determined by $\omega_{in}=\omega_{j}$ with $l=a/b$.
(1)When $b<b_{1}$ ($\omega_{in}<\omega_{1}$), the photon cannot propagate in the waveguide as the wavelength is larger than the cross section size. This is the so-called cut-off region.
(2) When $b_{1}<b<b_{2}$ ($\omega_{1}<\omega_{in}<\omega_{2}$), the principle of energy conservation leads to that only TM$_{11}$ mode can be propagated. This interval is the single-mode working region.
(3) When $b>b_{2}$ ($\omega_{in}>\omega_{2}$), multiple modes will appear simultaneously in the waveguide, which is called multi-mode working region. Again, it is the finite rectangular cross section of the waveguide that leads to the multiple TM modes with different cutoff frequencies.
\begin{figure}
\centering
\subfigure{\includegraphics[clip=true,height=4.3cm,width=7.0cm]{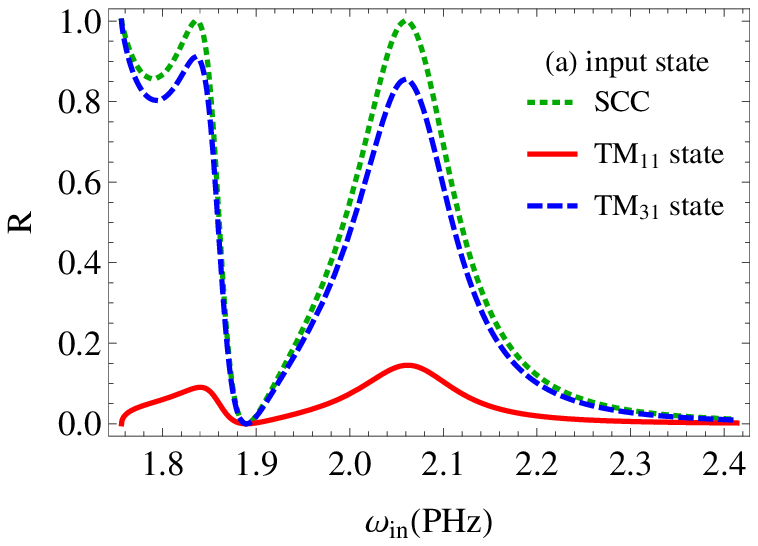}}
\subfigure{\includegraphics[clip=true,height=4.3cm,width=7.0cm]{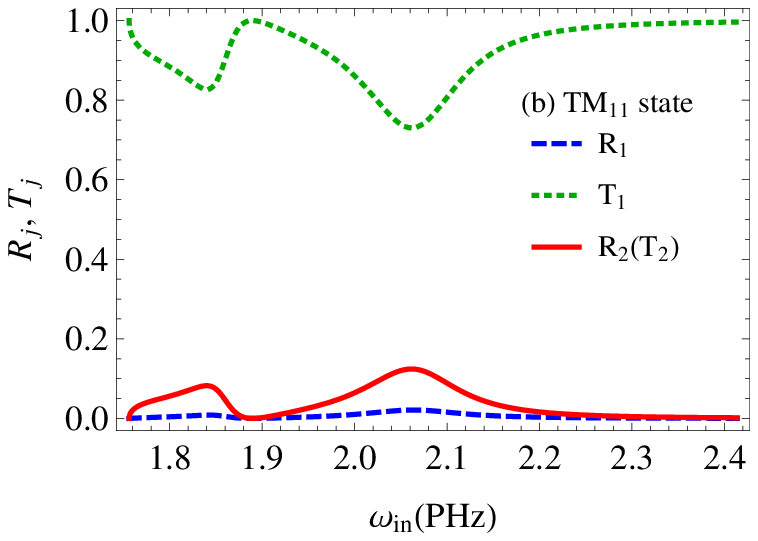}}
\subfigure{\includegraphics[clip=true,height=4.3cm,width=7.0cm]{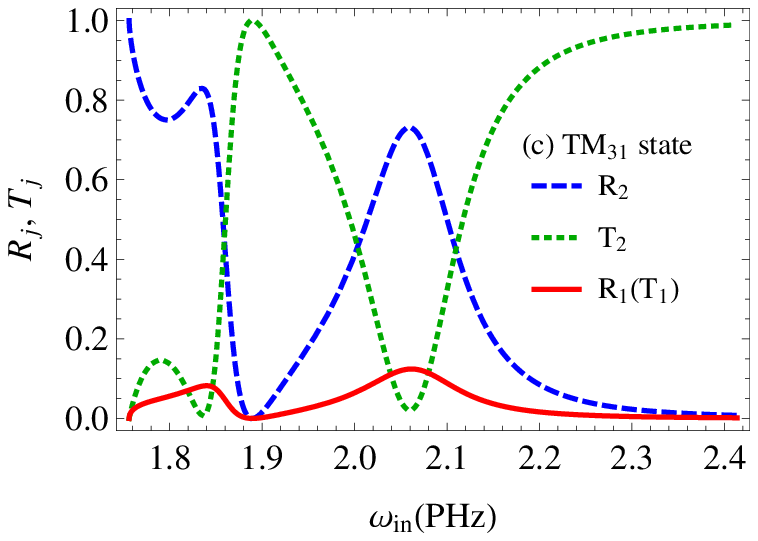}}
\caption{(Color online) Two-mode region: (a) The reflectivity $R$ against the input-photon energy $\omega_{2}<\omega_{in}<\omega_{3}$ with different input states. The dotted (green), solid (red) and dashed (blue) curves denote the input-photon state in the CSS, TM$_{11}$ and TM$_{31}$ sates, respectively.
The reflection $R_{i}$ and transition $T_{i}$ in the TM$_{11}$ mode (i=1) and TM$_{31}$ (i=2) against $\omega_{in}$ when the input sate is in the (b) TM$_{11}$-mode and (c) TM$_{31}$-mode. The size of cross section is $b=1.2$ um, $a=1.5b$. Other parameters are $\lambda_{1}=\lambda_{2}=0.05$, $\Omega_{1}=2$ PHz, $\Omega_{2}=1.8$ PHz.}\label{fig5}
\end{figure}

Next, we will carefully explore the multi-mode effect with different input states. When the input state is prepared in the CSS with $c_{j^{\prime }}\propto \omega _{j^{\prime }}/\sqrt{E^{2}-\omega _{_{j^{\prime }}}^{2}}$, the reflection spectrum in the multi-mode region is similar to that in the single-mode region. As discussed around Eq.~(\ref{eq-22}), all the results in subsection IV.~B almost apply to multi-mode case with input CSS, except that the frequency range changes. The dotted green line in Fig.~\ref{fig5} shows that, as long as the input state is prepared in the  CSS, both the perfect reflection and perfect transmission can be achieved. But when the input photon is in a single mode, the photon-transport characteristics in the multi-mode region are quite different from that in the single-mode region. For simplicity but without loss of generality, take the two-mode case with $\omega_{2}<\omega_{in}<\omega_{3}$ $(j_{\max}=2)$ for example.
The solid red and dashed blue lines in Fig.~\ref{fig5}(a)) show that, the perfect reflection is impossible when the input state is in either TM$_{11}$ or TM$_{31}$. But the perfect transmission can also be achieved under the condition of $y(\omega_{in})=0$, since the EIT mechanism still holds in the multi-mode region. All these results are consistent with the previous discussion on the perfect transmission and reflection conditions.

In order to further analyze the physical mechanism behind the multi-mode effect,
the reflection and transmission components in the $j$th TM mode can be derived by insetting the input state parameter $c_{j}=\delta_{nj}$ (assuming the input state in the $n$th TM mode) and Eq.~(\ref{eq-14}) into Eq.~(\ref{eq-15}),
\begin{eqnarray}
\label{eq-23}
R_{j}&=&\frac{\Lambda_{n}(\omega_{in})\Lambda_{j}(\omega_{in})}{\vert f(\omega_{in})\vert ^{2}}, \nonumber\\
T_{n}&=&\left|1-\frac{i\Lambda_{n}(\omega_{in})}{f(\omega_{in})}\right|^{2}, \ \ \ T_{j}=R_{j}\ (j\not=n),
\end{eqnarray}
where the $\Lambda_{j}(\omega_{in})$ stems from the transition from $|e_{i}\rangle$ to $|g\rangle$ $(i=1,2)$ with $\Gamma_{j}^{(i)}(\omega_{in})$ in Eq.~(\ref{eq-10}),
\begin{eqnarray}
\label{eq-24}
\Lambda_{j}(\omega_{in})&=&\Gamma_{j}^{(1)}(\omega_{in})(\omega_{in}-\Omega_{2})+\Gamma_{j}^{(2)}(\omega_{in})(\omega_{in}-\Omega_{1}).
\end{eqnarray}
Thus, the total reflectivity with the $n$th-mode input state can be obtained with $\Lambda(\omega_{in})=\sum_{j=1}^{j_{max}}\Lambda_{j}(\omega_{in})=\text{Im}[f(\omega_{in})]$,
\begin{eqnarray}
\label{eq-25}
R&=&1-T=\frac{\Lambda_{n}(\omega_{in})\Lambda(\omega_{in})}{\vert f(\omega_{in})\vert ^{2}}.
\end{eqnarray}

The physical picture of photon scattering in the multi-mode region can be described as follows: when a photon in the $n$th mode is incident, it may pass through directly, or it may be absorbed by the emitter with coupling strength $g_{n,k}^{(i)}$ and then re-radiated. It is worth noting that since the emitter is coupled to multiple TM modes,
the incident $n$th-mode photon inevitably enters other TM modes when the emitter transitions from $|e_{i}\rangle$ to $|g\rangle$. And because the re-radiated photon travels in random directions, the transmittance and reflectivity in other TM modes is exactly the same, i.e., $T_{j}=R_{j}$ ($j\not=n$). But for the $n$th mode, the transmission probability consists of directly pass through component and re-radiated component, which leads to $T_{n}\not=R_{n}$.
Also because of the coupling of emitter with multiple modes, even under the Fano resonance condition, the total reflectivity can only reach a peak value of $R_{\max}=\Lambda_{n}(\omega_{in})/\Lambda(\omega_{in})<1$, with the maximum reflectivity in the $j$th component $R_{j,\max}=\Lambda_{n}(\omega_{in})\Lambda_{j}(\omega_{in})/\Lambda(\omega_{in})^{2}$.
As the photon states density is larger in the higher mode for a given $\omega_{in}$, the reflectivity component in the high mode is therefore larger than that in the low modes ($R_{2}>R_{1}$). Last but not at least, the cutoff effect due to zero group velocity in the multi-mode region should be treated carefully. The photon will either be completely transmitted or completely reflected when the input-photon energy resonates with cut-off frequencies. It depends on whether the mode of incident photon and cut-off frequency is consistent, i.e., $R(\omega_{n})= 1$ and $R(\omega_{j\not=n})=0$.

Based on the above analysis, the photon transport characteristics in the multi-mode region can be summarized as follows. Only when the incident state is SCC, the photon can be completely reflected (under Fano resonance condition) and be completely transmitted (under EIT mechanism). But when the input state is in the single $n$th mode, due to the indirectly interaction between waveguide modes mediated by the emitter, the complete reflection vanishes and the photon inevitably enters other TM modes with equal $R_{j}=T_{j}$ ($j\not=n$). We can make further predictions that,  if the input $n$th mode is the lowest mode, the photon will be transmitted with great probability in a large input-frequency range, except for the small probability of reflection near the Fano resonance point. If the input $n$th mode is the highest mode that the energy allows, the reflection spectrum is similar to that in the single-mode region, except of the reduced reflectivity peak.

\section{\label{Sec:4} conclusion}
The single-photon scattering by a V-type three-level emitter in a rectangular waveguide has been studied for a larger range of input-photon energy. When the photon energy is in the single-mode regime, the perfect transmission and perfect reflection can both be achieved as long as the conditions equations $\text{Im}[f(\omega_{in})]=0$ and $\text{Re}[f(\omega_{in})]=0$ have solutions in the region of $\omega_{1}<\omega_{in}<\omega_{2}$. The physical mechanism of perfect transmission and reflection can be explained as EIT and Fano resonance, respectively.
Nevertheless, as the photon energy is increased to be greater than the second lowest cutoff frequency, more than one propagation mode is involved in scattering process, and the multi-mode effect due to the finite cross section should be considered. Although the perfect transmission still exists, the input-photon with a high energy can not be perfect reflected except that the input-state is prepared in the CSS with $c_{j^{\prime}}\propto\omega _{j^{\prime }}/\sqrt{E^{2}-\omega_{_{j^{\prime }}}^{2}}$. Noting that these conclusions cannot hold for the degeneracy three-level emitter or the two-level emitter cases, where only one complete reflection peak may appear and complete transmission will not occur. This exactly highlights the advantage of the non-degenerate three-level emitter. Besides, in all photon energy region, the perfect transmission frequency and the transmission spectral width can both be manipulated by adjusting the transition frequencies $\Omega_{i}$ and the coupling strengths $\lambda_{i}$.
Thus, our results may promote the development of single-photon devices with wide applicable frequency region, like optical switches and filters with narrow spectral width.
They also contribute to explore the effect of waveguide's boundary conditions on the photon transport characteristics.

\begin{acknowledgments}
This work was supported by the National Basic Research Program of China (Grants Nos. 2016YFA0301201 and 2014CB921403), the NSFC (Grants Nos. 11534002 and 11847010), the NSAF (Grants Nos. U1730449 and U1530401), the Hunan Provincial Natural Science Foundation of China (Grant No. 2019JJ50007), the Science Foundation of Hengyang Normal University (Grant No. 17D19), and the Open Foundation of Hunan Normal University (Grant No. QSQC1804).

\end{acknowledgments}.

\end{document}